\documentclass[a4paper]{jpconf}
\usepackage{graphicx}

\begin{document}
\title{Quantum-classical hybrid dynamics -- a summary}

\author{Hans-Thomas Elze}

\address{Dipartimento di Fisica ``Enrico Fermi'', Universit\`a di Pisa,  
Largo Pontecorvo 3, I-56127 Pisa, Italia}

\ead{elze@df.unipi.it}

\begin{abstract}
A summary of a recently proposed description of quantum-classical hybrids is 
presented, which concerns quantum and classical degrees of freedom of a composite object that interact directly with each other. This is based on notions of classical Hamiltonian mechanics suitably extended to quantum mechanics. 
\end{abstract}

\section{Introduction} 
 We briefly review an attempt to construct a theory that 
describes {\it quantum-classical hybrids}, consisting of quantum mechanical and 
classical objects that interact directly with each other \cite{me11,me12,me12b}. 
Hybrids might exist as a fundamentally different species of composite objects 
"out there",  with consequences for the range of applicability of quantum mechanics, 
or they may serve in an approximate description of certain complex quantum systems. 

The concept of quantum-classical  hybrids has recently been employed, in order to 
explore a hypothetical direct coupling of classical time machines to quantum 
objects \cite{me13}, while the application of hybrid dynamics to the experiment proposed by Penrose {\it et al.}  \cite{PenroseEtAl}, which is intended to test the 
existence of ``Schr\"odinger cat" 
states of spatially displaced macroscopic objects, and to the  manipulation of entanglement among two q-bits is presently under study \cite{tesi13}. 

\section{Quantum-classical hybrids}
\label{QuClHybr}  
The direct coupling of quantum mechanical (QM) and classical (CL) degrees of 
freedom -- {\it ``hybrid dynamics''} -- departs from quantum mechanics. Details of our 
approach can be found in Refs.\,\cite{me11,me12,me12b}, where also 
additional  references and discussion of related works are presented.  

Hybrid dynamics has been researched extensively for various reasons. -- 
For example, the Copenhagen interpretation of quantum mechanics entails the  
measurement problem which, together with the fact that quantum mechanics needs  
interpretation, in order to be operationally well defined, may indicate that 
it needs amendments. Such as   
a theory of the {\it dynamical} coexistence of  QM and CL objects.   
This should have  
impact on the measurement problem \cite{Sudarshan123} as well as on  
the description of the interaction between quantum matter and (possibly) classical 
spacetime \cite{BoucherTraschen}.  

Furthermore, it is of great practical interest to better 
understand QM-CL hybrids appearing in QM approximation 
schemes addressing many-body systems or interacting fields, 
which are naturally separable into QM and CL subsystems; for example, representing 
fast and slow degrees of freedom, mean fields and fluctuations, {\it etc.}  
(See also the article by Hu and Subasi in this volume for an informative discussion of some 
related issues \cite{HuSubasi13}.)   
 
Concerning the hypothetical emergence of quantum mechanics from some  
coarse-grained deterministic dynamics (see Refs.\,\cite{tHooft10,Elze09a,Adler} 
with numerous references to related work), the quantum-classical backreaction 
problem might  appear in new form, namely regarding the interplay of fluctuations 
among underlying deterministic 
and emergent QM degrees of freedom. Which can be 
rephrased succinctly as: {\it ``Can quantum mechanics be seeded?''}

\subsection{Consistency  requirements}
Thus, there is ample motivation for the numerous attempts to formulate a satisfactory  hybrid dynamics. Generally, however, they are deficient in one 
or another respect. 
Which has led to various no-go theorems, in view of the lengthy list of desirable properties or consistency requirements that ``{\it the}'' hybrid theory should 
fulfil, see, for example, 
Refs.\,\cite{CaroSalcedo99,DiosiGisinStrunz}:  
\begin{itemize} 
\item Conservation of energy. 
\item Conservation and positivity of probability. 
\item Separability of QM and CL subsystems in the absence of their interaction, 
recovering the correct QM and CL equations of motion, respectively. 
\item Consistent definitions of states and observables; existence of a Lie bracket structure 
on the algebra of observables that suitably generalizes  
Poisson and commutator brackets. 
\item Existence of canonical transformations generated by the observables; 
invariance of the classical sector under canonical transformations 
performed on the quantum sector only and {\it vice versa}. 
\item Existence of generalized Ehrenfest relations ({\it i.e.} the 
correspondence limit) which, 
for bilinearly coupled CL and QM oscillators,  
are to assume the form of the CL equations of motion  
(``Peres-Terno benchmark'' test \cite{PeresTerno}). 
\item `Free Will' \cite{Diosi11}. 
\item Locality. 
\item No-signalling. 
\item QM / CL symmetries and ensuing separability carry over to hybrids. 
\end{itemize} 

These issues have also been discussed for the hybrid ensemble theory of Hall and
Reginatto, which  does conform with the first six points listed  \cite{HallReginatto05,Hall08} but is in conflict with 
the last two \cite{me11,HallEtAll11,HallEtAll12}.  

We have proposed an alternative theory 
of hybrid dynamics based on notions of phase space \cite{me11}. 
This extends work by Heslot, demonstrating   
that quantum mechanics can entirely be rephrased in the language and 
formalism of classical analytical mechanics \cite{Heslot85}. 
Introducing unified notions 
of states on phase space, observables, canonical transformations, and a generalized 
quantum-classical Poisson bracket, this has led to the intrinsically 
linear hybrid theory to be summarized in the following, which allows to fulfil {\it all} of 
the above consistency requirements. 

Recently Buri\'c and collaborators have shown that the dynamical aspects of our 
proposal can indeed be derived for an all-quantum mechanical composite system by 
imposing constraints on fluctuations in one subsystem, followed by suitable coarse-graining \cite{Buric12,Buric12b,BuricDICE13}. 

\section{A representation of quantum mechanics in classical terms}
\label{QuMech} 
We recall that evolution of a {\it classical} object can be described in relation to its 
$2n$-dimensional phase space, its {\it state space}. A real-valued regular 
function on this space defines an {\it observable}, {\it i.e.}, a differentiable function 
on this smooth manifold.  

There always exist (local) systems of  
{\it canonical coordinates}, commonly denoted by $(x_k,p_k),\; k=1,\dots ,n$, 
such that the {\it Poisson bracket} of any pair of observables $f,g$ assumes 
the standard form (Darboux's theorem): 
\begin{equation}\label{PoissonBracket} 
\{ f,g \}\; =\; 
\sum_k\Big (\frac{\partial f}{\partial_{x_k}}\frac{\partial g}{\partial_{p_k}}
-\frac{\partial f}{\partial_{p_k}}\frac{\partial g}{\partial_{x_k}}\Big ) 
\;\;. \end{equation} 
This is consistent with $\{ x_k,p_l\}=\delta_{kl}$, $\{ x_k,x_l\} =\{ p_k,p_l\}=0,\; 
k,l=1,\dots ,n$, and has the properties defining a Lie bracket operation, 
mapping a pair of observables to an observable.  

General transformations ${\cal G}$ of the state space are restricted by compatibility with 
the Poisson bracket structure to so-called {\it canonical transformations}, which   
do not change physical properties of an object. They 
form a Lie group and it is sufficient 
to consider infinitesimal transformations. 
An {\it infinitesimal transformation} ${\cal G}$ is 
{\it canonical}, if and only if for any observable $f$ the map $f\rightarrow {\cal G}(f)$ 
is given by $f\rightarrow f'=f+\{ f,g\}\delta\alpha$, with some observable $g$, 
the so-called {\it generator} of ${\cal G}$, and $\delta\alpha$ an infinitesimal real 
parameter. -- Thus, for example, the canonical coordinates transform as follows: 
\begin{equation}\label{xpcan} 
x_k\;\rightarrow\;x_k'=x_k+\frac{\partial g}{\partial p_k}\delta\alpha 
\;\;,\;\;\;   
p_k\;\rightarrow\;p_k'=p_k-\frac{\partial g}{\partial x_k}\delta\alpha 
\;\;. \end{equation} 
This illustrates the fundamental relation between observables and generators 
of infinitesimal canonical transformations in classical Hamiltonian mechanics. 
\vskip 0.5cm 

Following Heslot's work, we learn that the previous analysis 
can be generalized and applied to quantum mechanics; this concerns 
the dynamical aspects as well as the notions of states, canonical transformations,  
observables, and measurements \cite{Heslot85}. 
\vskip 0.5cm 

The following derives from the fact that the {\it Schr\"odinger equation} and its 
adjoint are {\it Hamiltonian equations} following from an action principle 
\cite{me11}. We must add the {\it normalization condition}, 
${\cal C}:=\langle\Psi (t)|\Psi (t)\rangle\stackrel{!}{=}1\;$, 
for all state vectors $|\Psi\rangle$, which is    
essential for the probability interpretation of amplitudes;    
state vectors that differ by an unphysical constant phase are to be 
identified. Thus, the {\it QM state space} 
is formed by the rays of the underlying Hilbert space. 

\subsection{Oscillator representation}
\label{OscRepr}
A unitary transformation describes QM evolution, 
$|\Psi (t)\rangle =\hat U(t-t_0)|\Psi (t_0)\rangle$, 
with $U(t-t_0)=\exp [-i\hat H(t-t_0)/\hbar ]$, 
solving the Schr\"odinger equation. Thus, a stationary state, characterized by 
$\hat H|\phi_i\rangle =E_i|\phi_i\rangle$, with real energy eigenvalue $E_i$,  
performs a harmonic motion, {\it i.e.}, 
$|\psi_i(t)\rangle =\exp [-iE_i(t-t_0)/\hbar ]|\psi_i(t_0)\rangle
\equiv\exp [-iE_i(t-t_0)/\hbar ]|\phi_i\rangle$. We assume a denumerable set of 
such states. Following these observations, it is quite natural to introduce 
what we may call the {\it oscillator representation}. 

We expand state vectors with respect to a complete 
orthonormal basis, $\{ |\Phi_i\rangle\}$:  
\begin{equation}\label{oscillexp} 
|\Psi\rangle =\sum_i|\Phi_i\rangle (X_i+iP_i)/\sqrt{2\hbar} 
\;\;, \end{equation} 
where the time dependent coefficients are separated into  
real and imaginary parts, $X_i,P_i$. 
This expansion allows to   
evaluate the {\it Hamiltonian function} defined by    
${\cal H}:=\langle\Psi |\hat H|\Psi\rangle$: 
\begin{equation}\label{HamiltonianQM1} 
{\cal H}=\frac{1}{2\hbar}\sum_{i,j}\langle\Phi_i|\hat H|\Phi_j\rangle (X_i-iP_i)(X_j+iP_j) 
=:{\cal H}(X_i,P_i) 
\;\;. \end{equation} 
Choosing the set of energy eigenstates, $\{ |\phi_i\rangle\}$, 
as basis of the expansion, we obtain:     
\begin{equation}\label{HamiltonianQM2} 
{\cal H}(X_i,P_i)=\sum_i\frac{E_i}{2\hbar}(P_i^{\;2}+X_i^{\;2}) 
\;\;, \end{equation} 
hence the name {\it oscillator representation}. --  
Evaluating $|\dot\Psi\rangle =
\sum_i|\Phi_i\rangle (\dot X_i+i\dot P_i)/\sqrt{2\hbar}$ 
according to Hamilton's equations with ${\cal H}$ of 
Eq.\,(\ref{HamiltonianQM1}) or (\ref{HamiltonianQM2}), gives back the Schr\"odinger equation. -- 
Furthermore, the {\it normalization condition} becomes: 
\begin{equation}\label{oscillnormalization} 
{\cal C}(X_i,P_i)=\frac{1}{2\hbar}\sum_i(X_i^{\;2}+P_i^{\;2})\stackrel{!}{=}1 
\;\;. \end{equation} 
Thus, the vector with components given by 
$(X_i,P_i),\; i=1,\dots ,N$, is 
confined to the surface of a $2N$-dimensional sphere with radius $\sqrt{2\hbar}$, 
which presents a major difference to CL Hamiltonian mechanics. 

The $(X_i,P_i)$ may be considered 
as {\it canonical coordinates} for the state space of a 
QM object. Correspondingly, we introduce a {\it Poisson bracket}, 
cf. Eq.(\ref{PoissonBracket}), 
for any two {\it observables} on the {\it spherically compactified state space}, 
{\it i.e.} real-valued regular functions $F,G$ of the coordinates $(X_i,P_i)$:     
\begin{equation}\label{QMPoissonBracket} 
\{ F,G \}\; =\; 
\sum_i\Big (\frac{\partial F}{\partial_{X_i}}\frac{\partial G}{\partial_{P_i}}
-\frac{\partial F}{\partial_{P_i}}\frac{\partial G}{\partial_{X_i}}\Big ) 
\;\;. \end{equation} 
As usual, time evolution of an observable $O$ is generated by the Hamiltonian: 
$\mbox{d}O/\mbox{d}t=\partial_tO+\{ O,{\cal H} \}\;$.  
In particular, we find that the constraint of Eq.\,(\ref{oscillnormalization}) 
is conserved:  $\mbox{d}{\cal C}/\mbox{d}t=\{ {\cal C},{\cal H} \}=0\;$.

\subsection{Canonical transformations and quantum observables} 
\label{CanTransObserv}
In the following, we recall briefly the compatibility of the notion of 
observable introduced in passing above -- as in classical mechanics -- with the usual 
QM one. This can be demonstrated rigourously by the implementation of canonical transformations  and analysis of the role of observables as their generators.  
For details, see Refs.\,\cite{me11,me12,me12b,Heslot85}.   

The Hamiltonian function has been defined as observable in  
Eq.\,(\ref{HamiltonianQM1}), which relates it directly to the corresponding QM   
observable, namely the expectation of the self-adjoint Hamilton operator. 
This is indicative of the general structure with the following most important 
features.  
\subsubsection{Compatibility of unitary transformations and Poisson structure.} -- 
Classical canonical transformations are automorphisms of the state space 
which are compatible with the Poisson bracket. 
Automorphisms of the QM Hilbert space are implemented by 
unitary transformations. This implies a transformation of the canonical 
coordinates $(X_i,P_i)$ here. From this, one derives the invariance of   
the Poisson bracket defined in Eq.\,(\ref{QMPoissonBracket}) under unitary transformations. 
Consequently, the {\it unitary transformations on Hilbert space are canonical transformations on the $(X,P)$ state space}.   
\subsubsection{Self-adjoint operators as observables.} -- 
Any infinitesimal unitary transformation $\hat U$ can be generated by a self-adjoint operator 
$\hat G$, such that: $\hat U=1-(i/\hbar )\hat G\delta\alpha\;$,  
which leads to the QM relation between an observable and 
a self-adjoint operator. By a simple calculation, one obtains: 
\begin{equation}\label{XPcan} 
X_i\;\rightarrow\; X_i'=X_i+\frac{\partial \langle\Psi |\hat G|\Psi\rangle}
{\partial P_i}\delta\alpha 
\;\;,\;\;\; 
P_i\;\rightarrow\; P_i'=P_i-\frac{\partial \langle\Psi |\hat G|\Psi\rangle}
{\partial X_i}\delta\alpha 
\;\;. \end{equation} 
From these equations, the  
relation between an observable $G$, defined in analogy to classical mechanics 
(as above), and a self-adjoint 
operator $\hat G$ can be inferred: 
\begin{equation}\label{goperator} 
G(X_i,P_i)=\langle\Psi |\hat G|\Psi\rangle 
\;\;, \end{equation} 
{\it i.e.}, by comparison with the classical result.  
Hence, a {\it real-valued regular function $G$ of the state is an observable, if 
and only if there exists a self-adjoint operator $\hat G$ such that Eq.\,(\ref{goperator}) 
holds}. This implies that {\it all QM observables are quadratic forms} 
in the $X_i$'s and $P_i$'s, which are essentially fewer than in the corresponding CL 
case. Interacting QM-CL hybrids require additional discussion, cf. below. 
\subsubsection{Commutators as Poisson brackets.} -- 
From the relation (\ref{goperator}) between observables and self-adjoint operators and the  Poisson bracket (\ref{QMPoissonBracket}) one derives: 
\begin{equation}\label{QMPBComm} 
\{ F,G\}=\langle\Psi |\frac{1}{i\hbar}[\hat F,\hat G]|\Psi\rangle 
\;\;, \end{equation}  
with both sides of the equality considered as functions of the variables $X_i,P_i$ 
and with the commutator defined as usual. Hence,  
the {\it QM commutator is a Poisson bracket with respect to the $(X,P)$ state space} and 
relates the algebra of its observables to the algebra of self-adjoint operators.   
\vskip 0.5cm

In conclusion, quantum mechanics shares with classical mechanics an even dimensional 
state space, a Poisson structure, and a related algebra of observables. It  
differs essentially by a restricted set of observables and the requirements 
of phase invariance and normalization, which compactify the underlying Hilbert space 
to the complex projective space formed by its rays.  

\section{Quantum-classical Poisson bracket, hybrid states and their evolution}
\label{PoissonBrEvol}
The far-reaching parallel of classical and quantum mechanics, as we have seen,    
suggests to introduce a {\it generalized Poisson bracket} for QM-CL hybrids:  
\begin{eqnarray}\label{GenPoissonBracket} 
\{ A,B\}_\times &:=&\{ A,B\}_{\mbox{\scriptsize CL}}+\{ A,B\}_{\mbox{\scriptsize QM}}
\\ [1ex] \label{GenPoissonBracketdef} 
&:=&\sum_k\Big (\frac{\partial A}{\partial_{x_k}}\frac{\partial B}{\partial_{p_k}}
-\frac{\partial A}{\partial_{p_k}}\frac{\partial B}{\partial_{x_k}}\Big )+  
\sum_i\Big (\frac{\partial A}{\partial_{X_i}}\frac{\partial B}{\partial_{P_i}}
-\frac{\partial A}{\partial_{P_i}}\frac{\partial B}{\partial_{X_i}}\Big ) 
\;\;, \end{eqnarray} 
of any two observables $A,B$ defined on the Cartesian product of CL {\it and} QM 
state spaces. It shares the usual properties of a Poisson bracket. -- Note that due to the 
convention  introduced by Heslot \cite{Heslot85}, to which we adhered in 
Sect.~\ref{QuMech},  the QM variables $X_i,P_i$ have dimensions of 
(action$)^{1/2}$ and, consequently, no $\hbar$ appears in 
Eqs.\,(\ref{GenPoissonBracket})--(\ref{GenPoissonBracketdef}). At the expense 
of introducing appropriate rescalings, these variables could be made to have their 
usual dimensions and $\hbar$ to appear explicitly here. -- For the remainder  
of this article, instead, we choose units such that $\hbar\equiv 1$.
 
Let an {\it observable ``belong'' to the CL (QM) sector, if it is  
constant with respect to the canonical coordinates of the QM (CL) sector}. Then,   
the $\{\;,\;\}_\times$-bracket has the important properties: 
\begin{itemize} 
\item 
It reduces to the Poisson brackets introduced 
in Eqs.\,(\ref{PoissonBracket}) and (\ref{QMPoissonBracket}), respectively,   
for pairs of observables that belong {\it either} to the CL {\it or} the QM sector. 
\item 
It reduces to the appropriate one of the former brackets, 
if one of the observables belongs only to either one of the two sectors. 
\item 
It reflects the {\it separability} of CL and QM sectors, 
since $\{ A,B\}_\times =0$, if $A$ and $B$ belong to different sectors.  
\end{itemize} 
Hence, {\it if a canonical tranformation 
is performed on the QM (CL) sector only, then observables that belong to the 
CL (QM) sector remain invariant.}

\subsection{The hybrid phase space density}
The hybrid density $\rho$ for a self-adjoint density operator $\hat\rho$ in a given 
state $|\Psi\rangle$ is defined by  \cite{me11} :  
\begin{equation}\label{rhodens}
\rho (x_k,p_k;X_i,P_i):=\langle\Psi |\hat\rho (x_k,p_k)|\Psi\rangle
=\frac{1}{2}\sum_{i,j}\rho_{ij}(x_k,p_k)(X_i-iP_i)(X_j+iP_j)
\;\;, \end{equation} 
using  Eq.\,(\ref{oscillexp}) and   
$\rho_{ij}(x_k,p_k):=\langle\Phi_i|\hat\rho (x_k,p_k)|\Phi_j\rangle 
=\rho_{ji}^\ast (x_k,p_k)$. It describes a {\it QM-CL hybrid ensemble} 
by a real-valued, positive semi-definite, normalized, and possibly time dependent 
regular function on the Cartesian 
product state space canonically coordinated by $2(n+N)$-tuples $(x_k,p_k;X_i,P_i)$; 
the variables $x_k,p_k,\; k=1,\dots ,n$ and $X_i,P_i,\; i=1,\dots ,N$ are reserved 
for CL and QM sectors, respectively. 

Note that the relation between an observable in $(X,P)$-space 
and a self-adjoint operator, Eq.\,(\ref{goperator}), can be written as: 
$G(X_i,P_i)=\mbox{Tr}(|\Psi\rangle\langle\Psi |\hat G)$, with  
the QM pure state acting as one-dimensional projector here. Concerning 
the hybrid density $\rho$, 
we may use the representation of $\hat\rho$ in terms of its eigenstates,   
$\hat\rho =\sum_jw_j|j\rangle\langle j|$, to obtain: 
\begin{equation} \label{rhointerpr}
\rho (x_k,p_k;X_i,P_i)=\sum_jw_j(x_k,p_k)\mbox{Tr}(|\Psi\rangle\langle\Psi |j\rangle\langle j|) 
=\sum_jw_j(x_k,p_k)|\langle j|\Psi\rangle |^2  
\;\;,  \end{equation} 
with $0\leq w_j\leq 1$ and $\sum_j\int\Pi_l(\mbox{d}x_l\mbox{d}p_l)w_j(x_k,p_k)=1$. 
-- This shows  
that $\rho (x_k,p_k;X_i,P_i)$ is the {\it probability density to find in the hybrid 
ensemble the QM pure state} $|\Psi\rangle$, parametrized by $X_i,P_i$ through 
Eq.\,(\ref{oscillexp}), {\it together with  the CL state} given by a point in phase space, 
specified by coordinates $x_k,p_k$.  

Further remarks about   
superposition, pure/mixed, or separable/entangled QM states that may enter
the hybrid density can be found in Ref.\,\cite{me12,Buric12b}.  Generally, 
the simplest observable-like form of $\rho$ -- a bilinear function of QM 
``phase space'' variables $X_i,P_i$, as above -- has to be replaced for interacting hybrids,  
allowing for the form of an {\it almost-classical observable}; see Sect.~5.4 in  
\cite{me11} and a more detailed study discussing measurements in \cite{me12b}.  

In particular, the density $\rho$ evolves according to the hybrid 
Liouville equation (see the following subsection). This will lead to a solution $\rho (x_k,p_k;X_i,P_i;t)$ that is not necessarily bilinear in the QM variables $X_i,P_i$. 
However, we can (re)construct the self-adjoint density operator for the 
QM subsystem of the hybrid by: 
\begin{equation}\label{rhorec} 
\hat\rho(x_k,p_k;t)=
\Gamma_N^{-1}\int_{\delta S_{2N}(\sqrt 2)}\Pi_i(\mbox{d}X_i\mbox{d}P_i)\; 
\rho (x_k,p_k;X_i,P_i;t)|\Psi (X_i,P_i)\rangle\langle\Psi (X_i,P_i)| 
\;\;, \end{equation} 
i.e., conditioned on the state of the CL subsystem parametrized by 
$(x_k,p_k)$ and consistent with Eq.\,(\ref{rhodens});  the measure of integration 
here has been evaluated in detail in Ref.\,\cite{me11}. 

\subsection{Phase space evolution}
The {\it Liouville equation} 
for the evolution of hybrid ensembles \cite{me11} must be    
based on  the generalized Poisson bracket defined in 
Eqs.\,(\ref{GenPoissonBracket})--(\ref{GenPoissonBracketdef}) and 
Liouville's theorem. This leads us to: 
\begin{equation}\label{rhoevol} 
-\partial_t\rho = \{\rho ,{\cal H}_\Sigma\}_\times  
\;\;, \end{equation}
with ${\cal H}_\Sigma\equiv{\cal H}_\Sigma (x_k,p_k;X_i,P_i)$ and:  
\begin{equation}\label{HtotalInt} 
{\cal H}_\Sigma:={\cal H}_{\mbox{\scriptsize CL}}(x_k,p_k)
+{\cal H}_{\mbox{\scriptsize QM}}(X_i,P_i) 
+{\cal I}(x_k,p_k;X_i,P_i)   
\;\;, \end{equation} 
defining the relevant Hamiltonian function, including a hybrid interaction;  
${\cal H}_\Sigma$ is required to be an {\it observable}, in order to have 
a meaningful notion of energy. -- Note that {\it energy conservation} follows from 
$\{{\cal H}_\Sigma,{\cal H}_\Sigma\}_\times =0$.  

Finally, the Liouville equation describes a Hamiltonian flow, which implies that:  \\
$\bullet$
The normalization and positivity of the probability 
density $\rho$ are conserved in presence of 
a hybrid interaction; hence, its interpretation remains valid under hybrid evolution. 

\section{Conclusion} 
Besides constructing the QM-CL hybrid formalism, as outlined here, and showing how 
it conforms with the strong consistency requirements presented in Sec.~2.1, we earlier 
discussed  the possibility to have classical-environment induced decoherence, 
quantum-classical backreaction, a deviation from the Hall-Reginatto proposal 
\cite{HallReginatto05, Hall08} in 
presence of translation symmetry, and closure of the algebra of hybrid observables 
\cite{me11,me12b,Buric12}. 
Questions of locality, symmetry vs. separability, incorporation of 
 superposition, separable, and entangled QM states, and `Free Will' were considered 
in Ref.\,\cite{me12,Diosi11,Buric12b}, while an exotic application to (a class of)  
time machines coupled to quantum mechanical degrees of freedom has been shown 
in Ref.\,\cite{me13} and further studies of the Penrose {\it et al.} experiment, designed 
to study reduction (or not) of macroscopic Schr\"odinger cat states  \cite{PenroseEtAl},  and of entanglement control are underway \cite{tesi13}. 

\ack
It is a pleasure to thank N. Buri\'c and T. Skinner for discussions and correspondence.

\end{document}